\documentclass[final]{aipproc}
\layoutstyle{8x11double}

\usepackage{exscale}
\usepackage[intlimits]{amsmath}
\usepackage{amsmath}
\usepackage{amsfonts}
\usepackage{amssymb,amscd}                
\usepackage{array}
\usepackage{wrapfig}
\usepackage{bbm}
\usepackage{multirow}
\usepackage[x11names]{xcolor}
\usepackage{graphicx}
\usepackage[T1]{fontenc}
\usepackage{subfigure}

\usepackage[displaymath, tightpage]{preview}

\begin{document}

\title{Hadronic contribution to the muon g-2 from a Dyson-Schwinger perspective}

\classification{14.60.Ef\,\, 12.38.Lg\,\, 13.25.Cq}

\keywords{muon g-2, light-by-light scattering, Dyson-Schwinger equations}

\author{Tobias Goecke}{
  address={Institut f\"ur Kernphysik, 
 Technische Universit\"at Darmstadt, 
 Schlossgartenstra{\ss}e 9, 64289 Darmstadt, Germany}
}

\author{Christian S. Fischer}{
  address={Institut f\"ur Theoretische Physik, 
 Universit\"at Giessen, 35392 Giessen, Germany},
 altaddress={Gesellschaft f\"ur Schwerionenforschung mbH, 
 Planckstr. 1  D-64291 Darmstadt, Germany}
}

\author{Richard Williams}{
  address={Institut f\"ur Kernphysik, 
 Technische Universit\"at Darmstadt, 
 Schlossgartenstra{\ss}e 9, 64289 Darmstadt, Germany}
}

\begin{abstract}
A novel approach towards the hadronic contributions to the
anomalous magnetic moment of the muon $a_{\mu}$ is presented, namely the
Dyson-Schwinger equations of QCD. It has the advantage of being valid for
all momentum scales and has the potential to address off-shell amplitudes.
We present our first results for the pseudoscalar (PS) meson exchange 
and the quark loop contributions. The meson exchange ($\pi^0, \eta, \eta'$),
$a_\mu^{\textrm{LBL;PS}}=(84 \pm 13)\times 10^{-11}$,
is commensurate with previous calculations, while the quark loop contribution
$a_\mu^{\textrm{LBL;quarkloop}} = (107 \pm 48)\times 10^{-11}$,
is strongly enhanced by vertex dressing effects in the quark photon vertex.
Taken seriously this leads to the estimate of $a_\mu=116\,591\,865.0(96.6)\times 10^{-11}$,
giving a 1.9 $\sigma$ deviation between theory and experiment.
\end{abstract}

\maketitle

%%%%%%%%%%%%%%%%%%%%%%%%%%%%%%%%%%%%%%%%%%%%
%% MAINMATTER
%%%%%%%%%%%%%%%%%%%%%%%%%%%%%%%%%%%%%%%%%%%%
\section{introduction}

The anomalous magnetic moment $a_\mu = (g_\mu -2)/2$ of the muon is one of
the most precisely determined quantities in particle physics.
Experimental efforts at Brookhaven and theoretical efforts of the past ten years have
pinned $a_\mu$ down to the $10^{-11}$ level, leading to significant deviations
between theory \cite{Jegerlehner:2009ry} and experiment \cite{Bennett:2006fi} 
of $\simeq~3~\sigma$:
      \begin{align}
		\label{eqn:amuexperiment}
            \mbox{Experiment:} \,\,\,\,
			&116\,592\,089.0(63.0)\times 10^{-11}\;\;, 
		\\
		\label{eqn:amutheoretical}
            \mbox{\phantom{wwu}} \mbox{Theory:} \,\,\,\,
			&116\,591\,790.0(64.6)\times 10^{-11}\;\;.
      \end{align}
This discrepancy makes $a_\mu$ very interesting since it might
be taken as a hint towards physics beyond the standard model (SM). 
In order to confirm this hypothesis the uncertainties of both theory and
experiment have to be reduced even further.

\begin{figure}[t!]
	\includegraphics[width=0.27\columnwidth]{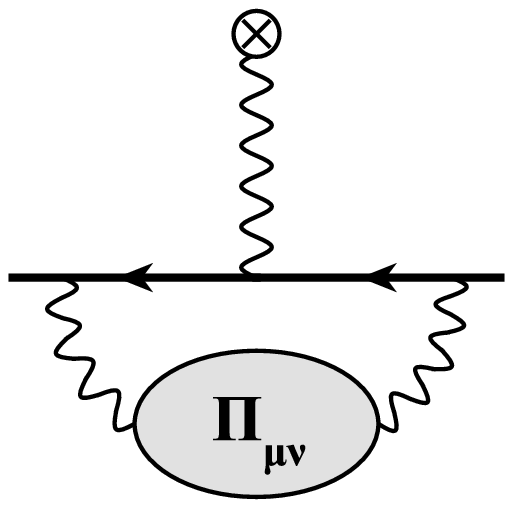}
			\hspace{0.08\columnwidth}
	\includegraphics[width=0.27\columnwidth]{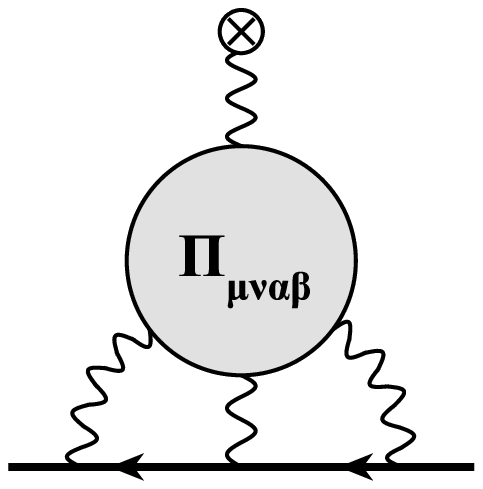}
      		\caption{Classes of corrections to the photon-muon vertex
	function: (left) Hadronic vacuum polarisation contribution;
	     	   (right) Hadronic light-by-light (LBL) scattering contribution.}
	\label{fig:hadroniclo}
\end{figure}

The theoretical error is dominated by hadronic contributions, \emph{e.g} diagrams
that involve QCD beyond perturbation theory (see \cite{Jegerlehner:2009ry}). 
The leading QCD contribution is given by the hadronic vacuum polarisation (HVP) 
insertion shown in Fig. \ref{fig:hadroniclo} (left).
%%%%%%%%%%%%%%%%%%%%%%%%%%%%%%%%%%%%%%%%%%%%%%%%%%%%%%%%%%%%%%%%%%%%%%%%%%%%%%%%%%
%%%%%%%%%%%%%%%%%%%%%%%%%%%%%%%%%%%%%%%%%%%%%%%%%%%%%%%%%%%%%%%%%%%%%%%%%%%%%%%%%%
This diagram also dominates the present error of the SM prediction.
However, since the HVP can be related to experimental data from
measurements of $e^+ e^- \rightarrow$ hadrons, a systematic reduction of the 
uncertainty will come with improved experimental input.

When this is realised, another contribution will dominate the theoretical error.
This is the hadronic light-by-light (LBL) scattering shown in Fig. \ref{fig:hadroniclo} (right).
For this contribution one must rely entirely on theory since no direct experimental
constraints are available. This contribution involves the hadronic photon
four-point function which is the central object in the following considerations.
This Green's function has been the subject of intensive investigation in
the past, mainly from the perspective of large-$N_c$ and chiral effective theory.
These suggest an ordering of diagrams that serve as an approximation towards
the full function \cite{deRafael:1993za}. The diagrams have been
calculated in the extended Nambu--Jona-Lasinio model (ENJL)
\cite{Bijnens:1995cc,Bijnens:2007pz} and the hidden local symmetry model
(HLS) \cite{Hayakawa:1995ps}. Later a refined analysis of the potentially
leading $\pi^0$ exchange contribution based on ideas of vector meson dominance
(VMD) has been started \cite{Knecht:2001qf, Melnikov:2003xd,Nyffeler:2009tw}.
Therein, the off-shell behaviour of the $\pi\rightarrow\gamma\gamma$ transition form factor
was considered. Recently, an analysis of LBL within the non local chiral
quark model (NL$\chi$QM) has been presented \cite{Dorokhov:2004ze}.\\
The hadronic LBL contribution to $a_\mu$ involves a two-fold
integration of the four point function (Fig. \ref{fig:hadroniclo} (right))
which makes it a two-scale problem. Thus a separation of hard and
soft scales is in general not straightforward.
We circumvent the process of matching an ultraviolet (UV) description
in terms of perturbative quarks and gluons and effective mesonic degrees
of freedom in the infrared (IR) by choosing a description that is entirely
based on the fundamental degrees of freedom of QCD, quarks and gluons.
In this non-perturbative approach, bound-states arise dynamically.
We rely on the Dyson-Schwinger equations (DSE)
which provide a non-perturbative means to obtain full
one particle-irreducible (1PI) Green's functions~\cite{Alkofer:2000wg}.
In particular we work in a truncation scheme defined in Ref.~\cite{Maris:1999nt} (see
the next section). Further details can be found in
\cite{Fischer:2010iz,longpaper} and extensions beyond this truncation can be found
in~\cite{Fischer:2007ze}.

In the present model the photon four point function may 
be defined as a resummation of an infinite subset of the leading order $1/N_c$
diagrams. In particular only planar diagrams without internal
quark lines are considered. Additionally, the Yang-Mills sector of QCD is truncated
to two-point functions so that only 'rainbow-ladder' gluon insertions
are taken into account. This leaves us with the expansion shown in Eq.~(\ref{fig:photon4pt_2})
%%%%%%%%%%%%%%%%%%%%%%%%%%%%%%%%%%%%%%%%%%%%%%%%%%%%%%%%%%%%%%%%%%%%%%%%%%%%%%%%%%
\begin{align}
	  \parbox{1.5cm}{\includegraphics[height=1.5cm]{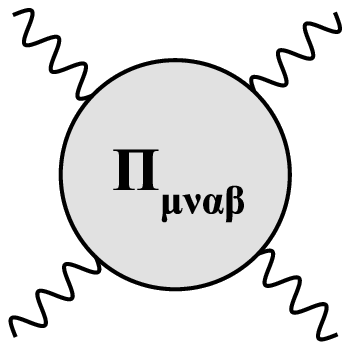}}
            \simeq
	    \parbox{1.5cm}{\includegraphics[height=1.5cm]{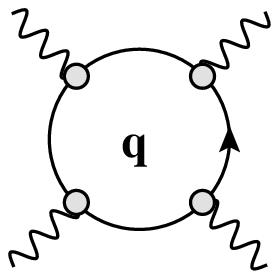}}
            +
	    \parbox{1.9cm}{\includegraphics[height=1.5cm]{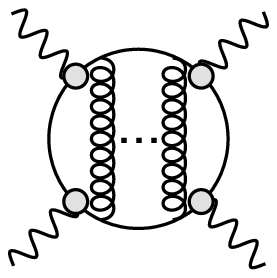}}
            \,+\,\ldots
\label{fig:photon4pt_2}
\end{align}
%%%%%%%%%%%%%%%%%%%%%%%%%%%%%%%%%%%%%%%%%%%%%%%%%%%%%%%%%%%%%%%%%%%%%%%%%%%%%%%%%%
where all quarks and quark-photon vertices are
fully dressed. It is well known that the second diagram
of this expansion includes various resonances and in particular pseudo-scalar
mesons. This gives rise to the pole 
approximation shown in Eq.~(\ref{fig:photon4pt})
%%%%%%%%%%%%%%%%%%%%%%%%%%%%%%%%%%%%%%%%%%%%%%%%%%%%%%%%%%%%%%%%%%%%%%%%%%%%%%%%%%
\begin{align}     
	  \parbox{1.5cm}{\includegraphics[height=1.5cm]{photon4ptfn.eps}}
            \simeq
	    \parbox{1.5cm}{\includegraphics[height=1.5cm]{photon4ptfn-qrkloop.eps}}
            +
	    \parbox{2.4cm}{\includegraphics[height=1.5cm]{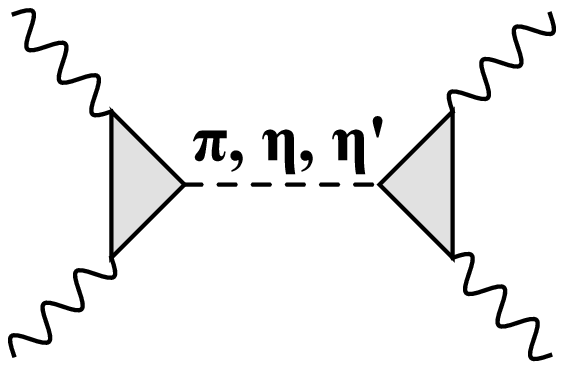}}
            \,+\,\ldots
\label{fig:photon4pt}
\end{align}			
%%%%%%%%%%%%%%%%%%%%%%%%%%%%%%%%%%%%%%%%%%%%%%%%%%%%%%%%%%%%%%%%%%%%%%%%%%%%%%%%%%
which becomes exact
on the respective meson mass shell. This leads us to a picture
that includes meson exchange terms in analogy to the $1/N_c$ picture.
While an evaluation of Eq.~(\ref{fig:photon4pt_2}) is currently underway
we present here results for Eq.~(\ref{fig:photon4pt}) as a starting point of our
investigations.

\section{Propagator and vertices within the DSE/BSE-approach}

\begin{figure}[b!]
      \includegraphics[width=0.80\columnwidth]{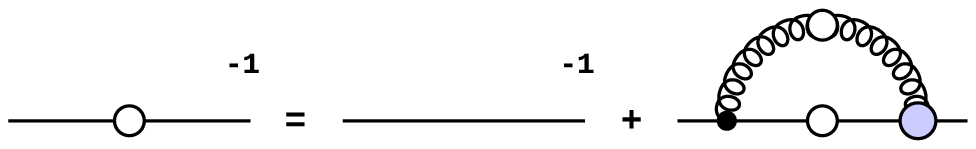}
      \caption{Dyson--Schwinger equation for the quark propagator.}\label{fig:quarkdse}
\end{figure}

\begin{figure}[b!]
      \includegraphics[width=0.45\columnwidth]{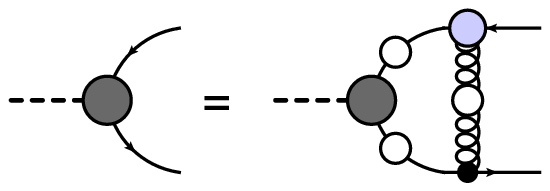}
      \caption{Bethe-Salpeter equation for quark-antiquark bound states.}
      \label{fig:bse}
\end{figure}

\begin{figure}[b!]
      \includegraphics[width=0.9\columnwidth]{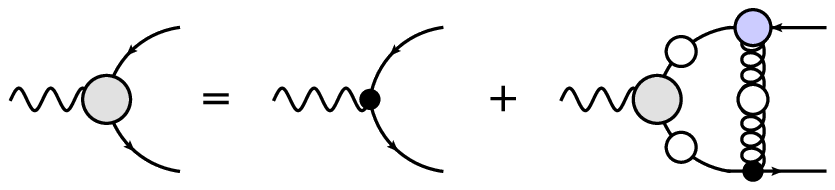}
      \caption{Inhomogeneous BS equation for the quark-photon vertex.}\label{fig:quarkphotondse}
\end{figure}

In the following we summarize our calculational scheme for LBL; details will be given 
elsewhere~\cite{Fischer:2010iz,longpaper}. The three objects
that we must calculate are the fully dressed quark propagator (Fig.~\ref{fig:quarkdse}), 
the Bethe-Salpeter-amplitude
(BSA, Fig.~\ref{fig:bse}) for pseudo-scalar (PS) mesons and
the quark-photon-vertex (Fig.~\ref{fig:quarkphotondse}). These are defined via 
self-consistent integral equations that require numerical solution.
%%%%%%%%%%%%%%%%%%%%%%%%%%%%%%%%%%%%%%%%%%%%%%%%%%%%%%%%%%%%%%%%%%%%%%%%%%%%%%%%%%
%%%%%%%%%%%%%%%%%%%%%%%%%%%%%%%%%%%%%%%%%%%%%%%%%%%%%%%%%%%%%%%%%%%%%%%%%%%%%%%%%%
%%%%%%%%%%%%%%%%%%%%%%%%%%%%%%%%%%%%%%%%%%%%%%%%%%%%%%%%%%%%%%%%%%%%%%%%%%%%%%%%%%
%%%%%%%%%%%%%%%%%%%%%%%%%%%%%%%%%%%%%%%%%%%%%%%%%%%%%%%%%%%%%%%%%%%%%%%%%%%%%%%%%%%
%%%%%%%%%%%%%%%%%%%%%%%%%%%%%%%%%%%%%%%%%%%%%%%%%%%%%%%%%%%%%%%%%%%%%%%%%%%%%%%%%%
%%%%%%%%%%%%%%%%%%%%%%%%%%%%%%%%%%%%%%%%%%%%%%%%%%%%%%%%%%%%%%%%%%%%%%%%%%%%%%%%%%

The gluon ($D_{\mu\nu}(k^2)$) and quark-gluon vertex are needed
to solve the quark DSE of Fig.~\ref{fig:quarkdse}.
In our truncation the vertex is considered to be projected onto one component
$\Gamma_\nu(p,q):=\Gamma^{YM}(k^2)\gamma_\nu$ giving the interaction
\begin{align}
  \parbox{2cm}{\includegraphics[width=0.2\columnwidth]{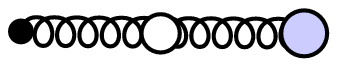}}
  & =  \gamma_\mu D_{\mu\nu}(k^2) \Gamma^{YM}(k^2)\gamma_\nu.
  \label{eqn:RLKernel}
\end{align}
The dressing $Z(k^2)$ of the landau gauge gluon $D_{\mu\nu}(k^2) =
(\delta_{\mu\nu}-k_\mu k_\nu/k^2)Z(k^2)/k^2$
is combined with the vertex dressing $\Gamma^{YM}(k^2)$ and modeled by 
a single dressing function that approaches one loop perturbative QCD (pQCD) in
the UV and contains a strongly enhanced IR part that induces dynamical chiral
symmetry breaking (D$\chi$SB) \cite{Maris:1999nt}.
The resultant quark contains pQCD above a few GeV and develops an
enhanced mass function of typically a few hundred MeV in the IR. In this way
perturbation theory is unified with the constituent quark picture with the quark
defined at all momentum scales.

Once the quark-DSE is solved, the quark, together with
(\ref{eqn:RLKernel}), forms an integral part of the remaining two equations 
for the three-point functions (Figs.~\ref{fig:bse} and \ref{fig:quarkphotondse}). 
This truncation respects the chiral properties of QCD such that low energy
theorems like the Gell-Mann-Oakes-Renner relation \cite{Maris:1998} are
fulfilled.
In this way observables such as bound-state masses, leptonic decay constant 
and form factors in the pseudo-Goldstone octet are reproduced on the percent level
while the accuracy is five to ten percent for vector mesons \cite{Maris:1999nt}.
These channels are the most important ones in the present calculation.

\begin{figure}[h]
%  \begin{center}
    \includegraphics[width=0.3 \columnwidth]{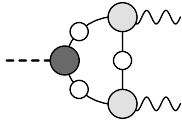}
%  \end{center}
  \caption{The $PS\rightarrow\gamma\gamma$ transition form factor.}
  \label{fig:quarktriangle}
\end{figure}

These building blocks now allow for a description
of QCD bound-states as solutions of the BSE (Fig.\ref{fig:bse}),
together with electromagnetic properties which are dependent on the 
quark-photon-vertex (Fig.\ref{fig:quarkphotondse}) that couples QCD to QED. 
We emphasize that the self-consistent quark-photon-vertex contains a dynamically
generated vector-meson bound state pole in the time-like region thus
encapsulating the ideas of vector meson dominance (VMD). More details can be found in
\cite{Maris:1999ta}. Consider as an example the
$\textrm{PS}\rightarrow\gamma\gamma$ transition form factor, shown in
impulse approximation in Fig.~\ref{fig:quarktriangle}.
This description fulfills the most important constraints for the case of
the pion, such the charge conservation in the correct asymptotic limits for high
photon virtualities. More information about the properties of this form
factor can be found in \cite{Maris:2002mz}.
In the present case, however, the pion is far away from its mass-shell
and thus an off-shell prescription is necessary (see \cite{longpaper} for details). 

With the quark, quark-photon vertex and $PS\rightarrow\gamma\gamma$
specified we now have all of the ingredients necessary to calculate the
photon four-point function and its expansion depicted in Eq.~(\ref{fig:photon4pt}).
\section{Results and Discussion}
The $\pi,~\eta,~\eta'$ exchange contributions amount to 
$a_\mu^{\textrm{LBL};\pi^0}=(57.5 \pm 6.9)\times 10^{-11}$,
$a_\mu^{\textrm{LBL};\eta} =(15.8 \pm 3.5)\times 10^{-11}$ and
$a_\mu^{\textrm{LBL};\eta'}=(11.0 \pm 2.4)\times 10^{-11}$ leading to
\begin{align}
   a_\mu^{\textrm{LBL;PS}}=(84.3 \pm 12.8)\times 10^{-11} \label{res:PS}.
\end{align}
The errors include numerical uncertainties as
well as estimates of the model dependence. The latter is determined by
consideration of meson observables in the respective channels.
The result given in Eq.~(\ref{res:PS}) compares well to previously
obtained values~\cite{Hayakawa:1995ps, Bijnens:1995cc, Knecht:2001qf}.

Due to numerical complexity, our results for the quark loop contribution do not 
yet include the full quark-photon vertex (Fig.~\ref{fig:photon4pt_2}). 
Instead we quote here results which include (a) bare vertices 
and (b) the leading part of the Ball-Chiu construction (1BC)~\cite{Ball:1980ay}:
	\begin{eqnarray}
		\begin{array}{lcc}
		a_\mu^{\textrm{LBL;quarkloop (bare vertex)}} &=&	(\phantom{0}61 \pm 2) \times 10^{-11}\\
		a_\mu^{\textrm{LBL;quarkloop (1BC)}}         &=& 	(107	\pm 2)		    \times 10^{-11}\\
		\end{array}\label{res:QL}
	\end{eqnarray}
The result for bare vertices in Eq. (\ref{res:QL}) is comparable to that 
found in constituent quark models. The 1BC vertex
strongly enhances the contribution to the muon anomaly.
This is in stark contrast to what has been found in other models
\cite{Bijnens:1995cc,Hayakawa:1995ps}. The difference of the 
two results in Eq. (\ref{res:QL}) determines the leading error
quoted above.
The inclusion of all non-perturbative covariant of the quark-photon
vertex is desirable, but is an extremely elaborate calculation and
deferred to future work.
At face value, this then leads to a revised estimate of 
the total $a_\mu=116\,591\,865.0(96.6)\times 10^{-11}$, which 
reduces the difference between theory and experiment to $\simeq1.9~\sigma$.

\section{Acknowledgments}
This work was supported 
by the Helmholtz-University Young Investigator Grant No.~VH-NG-332 and by the 
Helmholtz International Center for FAIR within the LOEWE program of the State of Hesse.

\bibliographystyle{aipproc}

\end{document}